\documentclass[twocolumn,prl,showpacs,byrevtex,superscriptaddress]{revtex4}
\usepackage{epsf,graphicx,amssymb}

\begin{document}
\title{Two-Dimensional Melting of a Crystal of Ferrofluid Spikes}
\author{Fran\c cois Boyer}
\affiliation{IUSTI, Universit\'e de Provence, CNRS (UMR 6595), 5 rue E. Fermi, 13 453 Marseille, France}
\author{Eric Falcon}
\affiliation{Laboratoire Mati\`ere et Syst\`emes Complexes (MSC), Universit\'e Paris Diderot, CNRS (UMR 7057) \\10 rue A. Domon \& L. Duquet, 75 013 Paris, France}
\date{\today}

\begin{abstract}  
We report the observation of the transition from an ordered solid-like  phase to a disordered liquid\-like phase of a lattice of spikes on a ferrofluid surface submitted to horizontal sinusoidal vibrations. The melting transition occurs for a critical spike displacement which is experimentally found to follow the Lindemann criterion, for two different lattice topologies (hexagonal and square) and over a wide range of lattice wavelengths. An intermediate hexatic-like phase between the solid and isotropic liquid phases is also observed and characterized by standard correlation functions. This dissipative out-of-equilibrium system exhibits strong similarities with 2D melting in solid-state physics.

\end{abstract}
\pacs{47.65.Cb, 64.70.D-, 05.70.Ln }

\maketitle

Melting of 3D crystals has been a substantial field of interest in condensed matter physics for over a century \cite{Einstein}. In 1910, Lindemann assumed that melting occurs when the thermal vibrations of atoms make them collide with each other \cite{Lindemann}. Then, it was suggested that the critical rms value of the atomic displacement at the melting transition is rather given by a constant fraction ($\simeq 10$~\%) of the interatomic distance \cite{Gilvarry}. This assumption, now known as the Lindemann criterion, has been succesfully used to predict melting temperatures of a wide range of 3D crystals \cite{Grimvall}. On the other hand, melting of a 2D solid is a less understood phenomenon since 2D lattices do not display true long-range translational order at finite temperatures \cite{Mermin}. Consequently, the Lindemann criterion was then long thought inapplicable in 2D \cite{Zheng}. A 2D melting theory driven by topological defects has then been developed by Kosterlitz, Thouless, Halperin, Nelson, and Young (KTHNY) \cite{KTHNY}. Numerical studies have then shown the validity of the Lindemann criterion in 2D \cite{Zheng}, and consistency with the KTHNY scenario \cite{Chen}. Although most experimental observations are consistent with this theory, it is quite difficult to establish unambiguously the existence of second-order solid-to-hexatic-to-liquid transitions \cite{Chaikin}. Consequently, the melting transition in 2D systems remains an active research field in solid-state physics \cite{Moucka}, and in numerous domains including thin colloidal suspensions \cite{Murray,Brunner02}, liquid films \cite{Klein}, vibrated granular monolayers \cite{Olafsen,Reis}, magnetic solid films~\cite{Sesha}, and vortex lattices in superconductors \cite{Scheidl}. 

A ferrofluid is a stable suspension of nanometric magnetic particles diluted in a carrier liquid which displays striking phenomena such as flows driven by a magnetic field gradient, magnetic levitation, labyrinthine and Rosensweig instabilities \cite{Rosen}. This latter occurs when a normal static magnetic field applied to a pool of ferrofluid exceeds a critical value: the flat free surface becomes unstable and a stationary hexagonal pattern of surface spikes grows. Following the pioneer work of Bragg {\it et al.} for the assemblage of soap bubbles \cite{Bragg2}, one can consider the Rosensweig spike lattice as a macroscopic analogous of a 2D crystalline structure and expect solid-like behaviors despite the complexity of the ferrohydrodynamic interaction between spikes and the dissipative nature of the lattice. An interesting feature of this system is that both the lattice wavelength and topology can be tuned by a single external control parameter (see below).

In this Letter, we report the first observation of a transition (melting) between an ordered phase (solid) and a disordered phase (liquid) of a lattice of spikes on the surface of a ferrofluid submitted to sinusoidal vibrations. We study which parameter controls the transition and whether a Lindemann criterion can be applied. We characterize structural changes across this transition using classical condensed matter physics concepts. Our system being dissipative with nonequilibrium steady states, the comparison of 2D melting in out-of-equilibrium systems and in equilibrium ones is of primary interest \cite{Olafsen,Reis}. 

\begin{figure}[t!]
\includegraphics[width=8.0cm]{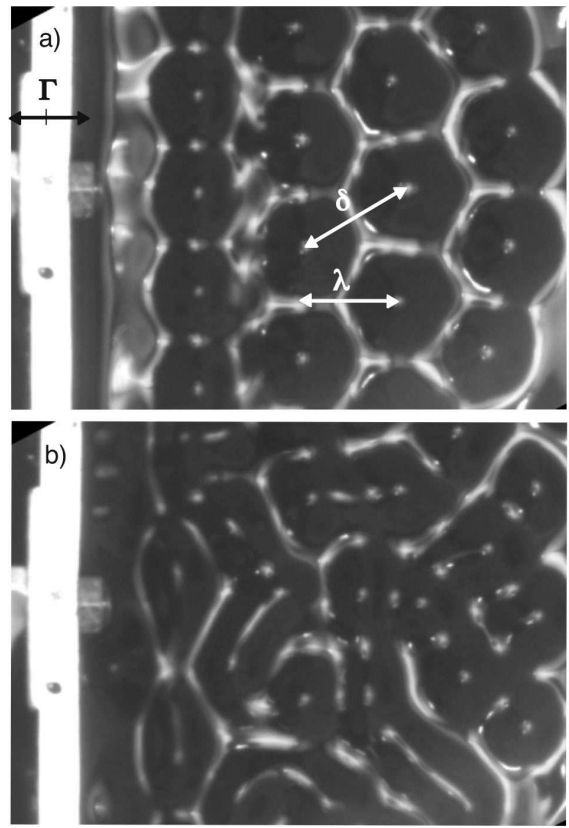} 
\caption{\label{fig01} Top views of the spike lattice on the surface of a ferrofluid for two different sinusoidal forcing amplitude: (a) hexagonal solid-like phase ($\Gamma = 3$ m.s$^{-2}$), $\delta \simeq 15.8$ mm, (b) liquid-like phase ($\Gamma = 20$ m.s$^{-2}$). The vibrating plate is shown in white, on the left-hand side. $f = 8$ Hz, $B^* = 1.2$.}
\end{figure}

The experimental setup has been described previously \cite{Boyer}. It consists of a container filled with a ferrofluid up to a depth $h=2$ cm. In order to discriminate any finite size or boundary condition effects, containers of different shapes and sizes are used: cylindrical containers i) 20 cm or ii) 12 cm in inner diameter, and iii) a rectangular container $13\times 9$ cm$^2$ sides. All containers are 4 cm depth. The ferrofluid used is a aqueous suspension of maghemite  particles \cite{Boyer}. Its properties are: density, $\rho=1324$ kg/m$^3$, surface tension, $\gamma=59\times 10^{-3}$ N/m, initial magnetic susceptibility, $\chi_i=0.69$, magnetic saturation $M_{sat}=16.9\times 10^{3}$ A/m, the viscosity being close to the water one. The container is placed in a vertical magnetic induction $B$ generated by two horizontal coaxial coils (up to 780G) \cite{Boyer}. A pattern of spikes on the surface of the ferrofluid is observed when $B$ is above a critical value $B_c=294\pm2$ G. This is close to the theoretical value of $292.3$ G computed as the threshold of the Rosensweig instability of our ferrofluid \cite{Boyer}. We denote the dimensionless magnetic induction $B^* = B/B_c$. Values $B^*>1$ will be used in the following. A hexagon-square transition occurs at a second threshold \cite{Gailitis,Abou}, and is observed here at $B^*=1.45$. For a fixed $B^*$, the lattice of spikes of ferrofluid is vibrated by means of the horizontal motion of a rectangular Teflon plate (plunging perpendicularly to the fluid at rest) that is driven sinusoidally by an electromagnetic vibration exciter at frequency $f$ and maximal displacement amplitude $a$ in the ranges $5 \leq f \leq 50$ Hz and $0.5 \leq a \leq 7$ mm, respectively. The vibrating plate acceleration is measured with an accelerometer. A high resolution camera located above the ferrofluid container allows us to measure the spike positions by calculating centers of the bright spots produced by reflections of light on the top of each spike (see Fig. \ref{fig01} for a typical snapshot). A particle tracking method is used to measure the spike displacement $d$ with respect to the vibrating plate displacement $a$. For the three containers used, $d$ is found to be a linear function of $a$ with a frequency-independent coefficient $\alpha = d/a$:  i) $\alpha = 0.77 \pm 0.04$; ii) $\alpha = 0.85 \pm 0.05$; iii) $\alpha = 0.65 \pm 0.05$. The slight variations of $\alpha$ with the container may be due to different relative positions of the vibrating plate to the front row of spikes. 
Since the vibrations are sinusoidal, the acceleration amplitude of the vibrating plate then reads $\Gamma = 4 \pi f^2a   =4 \pi f^2d/\alpha$.

Figure \ref{fig01} shows photos of the spike lattice on the ferrofluid surface for two different vibrating plate accelerations $\Gamma$, at a fixed $B^*$. At small $\Gamma$ (Fig. \ref{fig01}a), the lattice vibrates at the same frequency $f$ than the vibrating plate one: each spike displacement fluctuates about its equilib\-rium position as would do atoms of a crystal due to thermal motion. When $\Gamma$ is enough increased (at a constant $f$), the ordered lattice melts: ferrofluid spikes do not have a steady position anymore and rows of spikes slide past each other. At still higher $\Gamma$ (Fig. \ref{fig01}b), the spikes display highly disordered dynamics and two (or more) colliding spikes can fuse into one larger unstable spike.

\begin{figure}[h!]
\includegraphics[width=7.5cm]{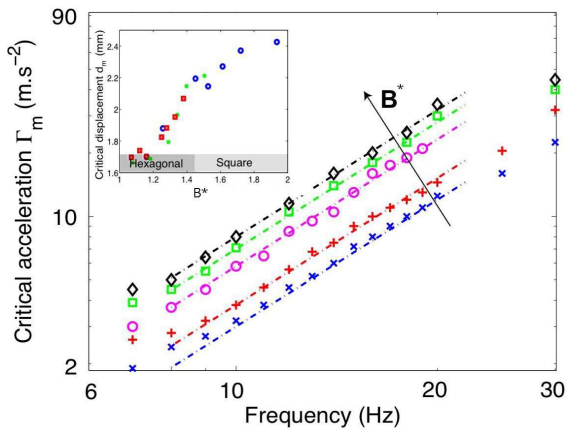} 
\caption{\label{fig02} (color online). Critical acceleration $\Gamma_{m}$ at the melting as a function of the vibration frequency $f$ for different dimensionless magnetic induction $B^*$: ($\times$) $1.09$, ($+$) $1.39$, ($\circ$) $1.52$, ($\square$) $1.77$ and ($\diamond$) $2.03$. Dashed lines have slopes 2 (constant displacement). Inset: Critical displacement $d_m$ as a function of $B^*$ for 3 different containers: ($\circ$) i, ($\times$) ii, and ($\square$) iii.}
\end{figure}

The onset of melting is defined once a spike has a motion larger than the lattice wavelength. It appears to be a sharp transition: the corresponding critical acceleration at the melting $\Gamma_{m}$ can be measured with an experimental error of less than 5\%. No hysteretic behavior is observed. Figure \ref{fig02} then shows the evolution of the critical acceleration $\Gamma_{m}$ as a function of the vibration frequency $f$ for 5 different applied magnetic inductions. For $8 \leq f \leq20$ Hz, $\Gamma_{m}$ is found to follow a power law with a scaling exponent $2 \pm 0.1$. This result means that along the melting transition, the quantity $\Gamma_m/f^2$ is a constant. 
Since the forcing is sinusoidal, this means that the amplitude of the spike displacement at melting, $d_m$, is the relevant parameter for the transition:
\begin{equation}
 d_m =   \frac{\alpha \Gamma_m }{4 \pi f^2}.
 \label{dm}
 \end{equation}
For each $B^*$, the value of $d_m$ is extracted, using Eq.\ (\ref{dm}), from the ordinate intercept of the slope of each log-log plot in Fig. \ref{fig02}. The inset of Fig. \ref{fig02} then shows the dependence of the critical displacement $d_m$ (rms value) with the dimensionless magnetic induction $B^*$ for different containers. $d_m$ is found to increase with $B^*$ from 1.6 mm to 2.6 mm. Note that these values are one order of magnitude lower than the typical wavelength between spikes $\lambda_c = 2 \pi \sqrt{\gamma/(\rho g)} \simeq 13.4$ mm at the Rosensweig instability threshold \cite{Rosen}.
 
\begin{figure}[t!]
\includegraphics[width=7.5cm]{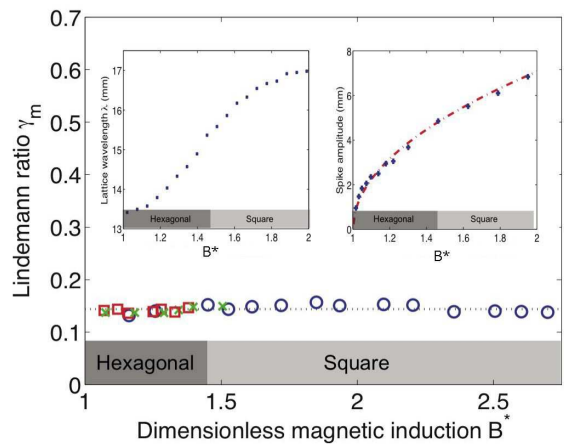} 
\caption{\label{fig03} (color online). Lindemann ratio $\gamma_m$ as a function of the dimensionless magnetic induction $B^*$ measured for 3 different containers: ($\circ$) i, ($\times$) ii, and ($\square$) iii. The dotted horizontal line is $\gamma_m = 0.14$. Left inset: Lattice wavelength, $\lambda$, as a function of $B^*$. Right inset: spike amplitude, $h$, as a function of $B^*$. $\sqrt{B^*-1}$ fit (dashed line).}
\end{figure}

In order to establish a criterion for the melting transition, both the lattice wavelength, $\lambda$, and the spike height, $h$, are measured to be compared to the spike displacement at melting, $d_m$. The amplitude $h$ of the ferrofluid spike is measured by means of a capacitive wire gauge \cite{Boyer}. Right inset of Fig. \ref{fig03} shows $h$ as a function of $B^*$: data can be well fitted by $\sqrt{B^*-1}$ (see dashed line) in good agreement with theory \cite{Gailitis} and a previous observation \cite{Bacri}. The melting displacement, $d_m(B^*)$ (inset of Fig. \ref{fig02}), clearly does not follow the same law than the spike amplitude one, $h(B^*)$ (right inset of Fig. \ref{fig03}), both qualitatively and quantitatively. This means that the melting mechanism is not induced by surface waves higher than spike height. For a fixed $B^*$, the wavelength $\lambda$ is determined by measuring the average spike-to-spike distance $\delta$ on experimental frames (see Fig.\ \ref{fig01}). From simple geometrical considerations, one has $\lambda = (\sqrt{3}/2) \delta$ for the hexagonal pattern ($1< B^*<1.45$), and $\lambda = \delta$ for the square pattern ($B^*>1.45$). Left inset of Fig. \ref{fig03} then shows $\lambda$ as a function of $B^*$. $\lambda$ is found to increase with $B^*$ as previously reported \cite{Abou}. 
Let us now define an analog of the Lindemann ratio in solid-state physics,
 \begin{equation}
\gamma_m \equiv \frac{d_m(B^*)}{\lambda(B^*)}\ {\rm ,}
 \end{equation}
that is the ratio between the rms value of the spike displacement, $d_m$, at the melting transition and the lattice wavelength $\lambda$ at a fixed $B^*$. As shown in Fig. \ref{fig03}, $\gamma_m$ is found to be independent of $B^*$, even at the hexagon-square transition. This is the first experimental observation of the Lindemann criterion for the 2D melting transition of a crystal of ferrofluid spikes. The melting Lindemann ratio $\gamma_m$ is found to be equal to $0.14 \pm 0.02$ which is in the range of 3D crystalline solid values (0.1 - 0.2) \cite{Gilvarry} and close to the reported value in 2D granular fluids (0.15) \cite{Reis}. Note that defining the Lindemann ratio with respect to the lattice wavelength, $\lambda$, instead of the spike-to-spike distance, $\delta$, leads to a single value for both lattice topologies (hexagonal and square).

\begin{figure}[t!]
\includegraphics[width=7.5cm]{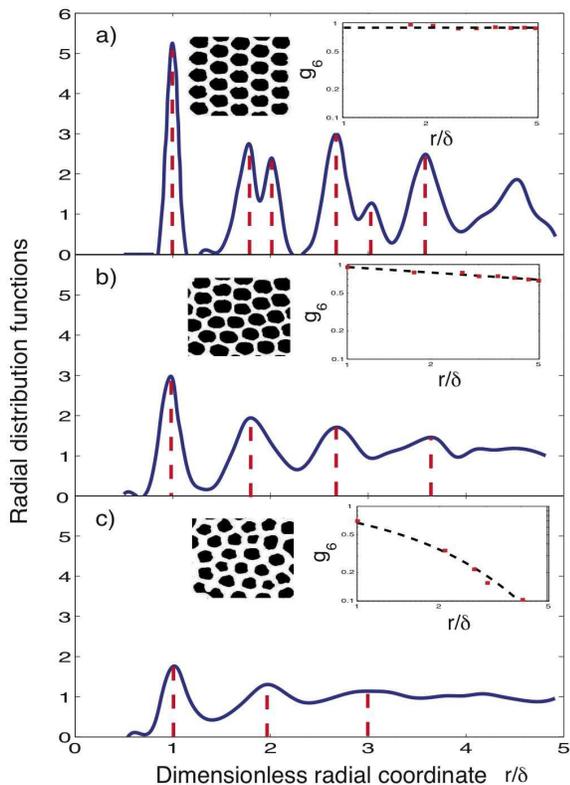} 
\caption{\label{fig04} (color online). Radial distribution functions $g(r/\delta)$ for different forcing: (a) $\varepsilon = -0.4$, (b) $\varepsilon = 0$, (c) $\varepsilon = 1$. $B^*= 1.1$. $f = 10$ Hz. $\delta= 15.5$ mm. Dashed lines show predicted maximum positions of $g(r)$ (see text). Right insets:  Log-log orientational correlation functions $g_6(r/\delta)$. Dashed lines are (a) $0.89$, (b) $(r/\delta)^{-0.20}$, (c) $\exp[-0.65(r/\delta)]$. Left insets: typical thresholded photos of the ferrofluid surface.}
\end{figure}

To characterize structural changes along the melting transition at a fixed $B^*$, one computes standard positional and orientational correlation functions, respectively, $g(r)\equiv \langle  n(r') n(r+r') \rangle/ \langle  n(r') \rangle^2$, where $n$ is the spike density at a distance $r$ from a reference (brackets are an average over the spatial variable $r'$),  and $g_6(r) \equiv \langle \Psi^*_6(r') \Psi_6(r+r') \rangle/g(r)$ where the star denotes the complex conjugate and $\Psi_6(r_j) \equiv \langle \exp(i6\theta_{jk}) \rangle_k$ with $\theta_{jk}$ the orientation angle of the bond between the centers of the spike $j$ and of the neighbouring spike $k$ \cite{Chaikin}.  These functions are averaged over 200 frames leading to an error bar of 2\%. Figure \ref{fig04} shows experimental curves of the radial distribution functions $g(r)$ for different values of the dimensionless forcing parameter $\varepsilon = (\gamma - \gamma_m)/\gamma_m$ where $\gamma \equiv d/\lambda$. For $\varepsilon < 0$ (before the melting), $g(r/\delta)$ displays characteristic features of a hexagonal structure: the first spikes positions are in very good agreement with values predicted from simple geometrical calculations $r/\delta=1$, $\sqrt{3}$, $2$, $\sqrt{7}$, $3$ and $\sqrt{12}$ (see dashed lines in Fig.\ \ref{fig04}a). For $\varepsilon = 0$, the positional order is clearly short-ranged although some of the characteristic spikes of the hexagonal lattice remain visible: $r/\delta=1$, $\sqrt{3}$, $\sqrt{7}$ and $\sqrt{12}$ (see dashed lines in Fig.\ \ref{fig04}b). As $\varepsilon$ is further increased, the positional-order range becomes shorter, and for  $\varepsilon = 1$, only the characteristic spikes of an isotropic liquid phase are observed: $r/\delta=1$, $2$ and $3$ (see dashed lines in Fig.\ \ref{fig04}c). These typical structural changes show strong similarities with the ones reported during numerical simulation of hard-disk fluid when the 2D solid-liquid phase transition is approached \cite{Moucka}. Our results are also consistent with the KTHNY theory in solid physics which predicts the existence of a hexatic phase between the crystalline and the liquid phases, characterized by a long-range orientational order (algebraic decay) and a short-range positional order (exponential decay) \cite{KTHNY}. Indeed, right-hand side insets of Fig.\ \ref{fig04} show the orientational correlation function, $g_6(r)$. Just above the melting transition ($\varepsilon \gtrsim 0$), the long-range orientational order (well fitted by an algebraic decay) is preserved as required for a hexatic phase (see Fig.\ \ref{fig04}b). Note that to utterly discriminate it from an exponential decay, a larger number of lattice periods should be necessary \cite{Olafsen}. For $\varepsilon = 0$, one has $g_6(r)\sim r^{-0.20}$ in agreement with KTHNY scenario as the power-law exponent is predicted to continuously decrease in the hexatic phase down to $-0.25$ at the hexatic-liquid transition \cite{KTHNY}.  As $\varepsilon$ is further increased, the decay rate increases strongly, and for $\varepsilon = 1$, the orientational order is short-ranged (well fitted by an exponential decay) as expected for a liquid phase. 

Such an intermediate hexatic phase between the crystal and liquid ones has been also reported in 2D melting of colloidal crystals with absolutely calibrated interaction in good agreement with the KTHNY theory \cite{Murray}. Note that the 2D melting of these systems can depend on the interaction with their carrier substrate \cite{Brunner02}. 
Our results put forward that both the Lindemann criterion and the KTHNY scenario are applicable to a more complex 2D system such as a crystal of ferrofluid spikes with complex ferrohydrodynamic interaction. Such a continuous solid-liquid transition, via a hexatic phase, also strongly differs of the first-order transition in 3D systems. Finally, our work emphasizes the analogy between 2D melting of equilibrium systems and out-of-equilibrium steady state ones. Such a correspondence has been observed in a 2D granular fluid \cite{Olafsen,Reis}, where spatial homogeneity of the energy injection within the system is underlined to be the main ingredient for these similarities with equilibrium dynamics \cite{Reis}. Our results suggest that equilibrium-like properties can be observed even though energy injection is inhomogeneous. This should deserve more studies to have a complete description of the 2D melting transition of a dissipative crystal. 

\begin{acknowledgments}
We thank J.-C. Bacri and A. Cebers for fruitfull discussion, A. Lantheaume, and C. Laroche for technical assistance. This work has been supported by ANR Turbonde BLAN07-3-197846.
\end{acknowledgments}

\end{document}